# OVERVIEW OF THE EXPERIMENTAL PHYSICS AND INDUSTRIAL CONTROL SYSTEM (EPICS) CHANNEL ARCHIVER

K.U. Kasemir, L.R. Dalesio, LANL, Los Alamos, NM 87544, USA

Abstract

The Channel Archiver[1] has been operational for more than two years at Los Alamos National Laboratory and other sites. This paper introduces the available components (data sampling engine, viewers, scripting interface, HTTP/CGI integration and data management), presents updated performance measurements [2] and reviews operational experience with the Channel Archiver.

## 1 INTRODUCTION

The Channel Archiver is a generic periodic sampling toolset for EPICS. Using the EPICS Channel Access (CA) network protocol[3], it can collect real-time data from any CA server on the network. In [1], two fundamentally different modes of archiving were discussed: Periodic sampling for historical data logging as opposed to taking a snapshot of data across many channels at a specific point in time for saving and restoring operational setpoints. The Channel Archiver implements the former and currently provides these tools: A sampling engine, command line data maintenance and export tools, a CGI export tool, graphical data browsers for Microsoft Win32 and Unix operating systems, and a scripting interface.

## 2 TOOLSET DESCRIPTION

Unless explicitly mentioned, the following programs are available for Win32 (Microsoft Windows 9x, NT, 2000) and Unix (Linux, Solaris and others).

The *ArchiveEngine* collects data from CA servers. A channel can be scanned at a fixed period or on change. Channels can be combined into a flat list of groups, archiving of a group can be suspended depending on the value of designated channels. A common application is to disable the channel group for a specific subsystem while it is inactive, e.g. power supply readbacks are not sampled while the device is off. The engine stores the full data set offered by CA: values, time stamps, status information as well as the engineering unit names, display-, control- and alarm limits. Special status values indicate disconnected channels or archiver shutdown. The ArchiveEngine has a built-in HTTP server for distributed and system independent access to status and configuration. Any web browser can be used to check what channels are connected or to modify their configuration.

*WinBrowser* is a data browser programs for Win32 with a graphical user interface. The user can list or filter channel names in the archive and plot selected channels with interactive zooming and panning. For further evaluation, data can be exported into various file formats. The generic Unix archive browser *Xarr* offers similar functionality. It is written by Chris Larrieu, TJNAF, as is the CA strip-chart client *StripTool*, which can be configured to access archived data.

*CGIExport* is a plug-in for web servers, providing archive access to any user with a web browser. It offers channel information, online plots and download of various file formats. The *ArchiveExport* command line tool can create the same assortment of data files; *ArchiveManager* is a command line tool for maintaining the binary data files.

The Channel Archiver Scripting Interface *CASI* provides users of the tcl, Python and Perl scripting languages[5-7] with access to the data. Scripting is a powerful and rapid development technique for data analysis that goes beyond simple retrieval. Scripts can also periodically update web pages or generate arbitrary reports.

All tools use a C++ Channel Archiver interface library. Based on the iterator model[4], this application programmer interface (API) provides abstract archive, channel and value classes. The current implementation supports both a single binary archive as well as reading from multiple binary archives. Additional formats can be added to this API, so that all tools could be adapted to an in-house data storage format. Similarly, a new export format can be added.

## 3 DATA EXPORT OPTIONS

The ArchiveEngine stores all samples as provided by the CA server, including the original time stamps. While this preserves the full nano-second resolution, it often prohibits easy correlations: Data from different channels will rarely carry matching time stamps. Even channels from the same scan process on the same machine are processed sequentially by the CPU, resulting in slightly different time stamps at nanosecond resolution. Therefore, the WinBrowser, CGIExport and ArchiveExport programs offer simple staircase and linear interpolation for correlating data from different channels. They produce files for spreadsheet programs, GNUPlot or Matlab.

Instead of adding more interpolation or plotting options to the ChannelArchiver tools, recent emphasis has been on support for existing commercial data analysis tools, namely the Matlab programming language[8]. It is extremely powerful for correlating, analyzing and visualizing data. The archiver sources include several example scripts for dealing with archived data in Matlab.

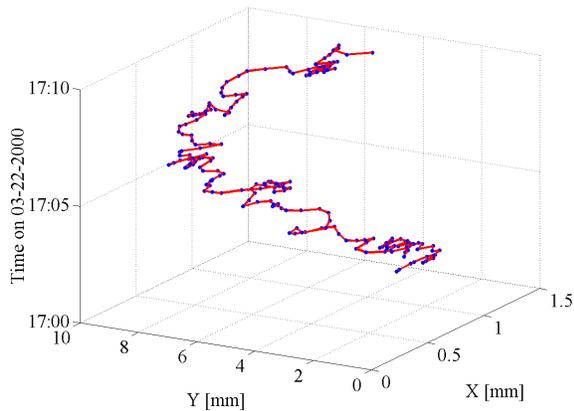

Figure 1: Matlab example for graphing a trajectory over time, assuming two archived process variables reflect the x respectively y position.

## 4 BINARY DATA FORMAT

We assume that analysis will concentrate on the last weeks if not hours of operation in most cases. The binary data storage format of the Channel Archiver is therefore optimized to support the addition of new samples to the end of the archive and retrieval of recent samples.

The binary archive data format uses a directory file and several data files. The combination of all these constitutes one "archive" to the user. The directory file is a disk-based hash table of channel names, including pointers to the first and last data buffer for each channel. The data files contain double-linked lists of sample buffers as well as control information buffers. Initially, each data file contains one buffer per channel. Whenever filled, new data buffers are appended to the same file until a configurable time limit is reached, which leads to the creation of a new data file (default: one file per day). Each value is saved with its time stamp and status. Additional control information (display limits, units, …) is written whenever it changes, at least once per data file.

The hash table quickly translates channel names into pointers to the last value buffer, facilitating the quick addition of new samples as well as the retrieval of the most recent values. In comparison, this format allowed writing around 15000 values per second on a Win32 PC where a commercial relational database package (Microsoft Access) could store only about 2400 values per second.

The ArchiveManager can list raw values, copy and delete channel names, and create new archives by copying a selected range of channels in time from an existing archive into a new instance. While copying, it skips invalid entries and compresses the archive files by creating a reduced number of channel buffers.

## 5 PERFORMANCE

The following tests apply to a RedHat 7.0 Linux PC with a RAID-5 disk array, 850 MHz CPU and 256 MB RAM. A C++ benchmark program for the Channel Archiver library can write more than 22000 values per second on this machine, indicative of the hard disk performance. An MVME2100 PPC running EPICS 3.13.3 under vxWorks 5.4 provided test data by serving 1000 channels, each generating ramping signals at 10 Hz.

The Archive Engine could successfully store the resulting 10000 values per second when configured to write to disk every 10 seconds. The CPU load was at 50%, rising to 100% every 10 seconds. A CASI perl script verified that no data was lost by comparing the archived values and time stamps to the expected ramping behavior. This script can test 21000 values per second.

A simple readout of all the values without further checking or output formatting can read over 40000 values per second. In a more realistic example, ArchiveExport produces around 600 lines of spreadsheet-formatted output per second for 50 channels, including linear interpolation onto 10Hz time stamps, which corresponds to handling 30000 values per second.

The CGIExport Web interface to the archived data offers similar throughput for spreadsheet generation, although the network connection to the web server can introduce unpredictable delays. For producing online plots, the CGIExport tool writes a data file on the server disk and then launches GNUPlot to create an image file of the actual plot. The overall time for this two-step process used to be about twice the spreadsheet creation time. A recent modification by Thomas Birke (BESSY) allows CGIExport to determine the minimum and maximum values of the original data for a number of "bins" that matches the width of the plot image, resulting in a significantly reduced data file and increased plotting speed. As a result, a plot of 50 channels that produced a data file of 423000 lines and took 30 seconds to complete will now finish in 4 seconds based on a 1500 line data file.

A common operation is finding a channel by name and retrieving the most recent values. For a one-month archive with 1400 channels, 770MB in 31 data files, a perl script can initially handle 60 name lookups per second, including the retrieval of the last 50 values for each channel. When run again, the script will perform 500 lookups per second because the operating system cashes the disk access. Within 30 minutes, the ArchiveManager copies this archive into a new archive of 480 MB in a single data file.

## 6 OPERATIONAL EXPERIENCE

Several EPICS sites are evaluating the Channel Archiver, including DESY, BESSY, TJNAF, KEK and SLAC. At Los Alamos, the LEDA project[2] has been using it since the first version became available. The main archive engine, running under Linux, is restarted monthly, collecting up to one CD-ROM (650 MB) of data in this period from around 1400 channels every 30 seconds. Data since July 1999 is kept online, totaling around 8 GB of disk space. Several subsystems each use a separate archive engine, and a designated "fast" archiver is running only when manually started during operations, so far requiring about 3 GB.

The Swiss Paul Scherrer Institute (PSI) is a more active archive user with Linux (RedHat 6.x) PCs dedicated to archiving and web serving[9]. Last year, the archiver collected about one CD-ROM (650 MB) of data per week. This year, a graded approach is used: A long-term archiver (~1030 channels every 10 minutes) runs continuously. A middle-term archiver (~8200 channels every 60 seconds) switches between two storage directories every 15 days, overwriting the previous data, keeping a record of the last month. A short-term archiver (~1500 channels, various frequencies and change monitors) switches between two directories, keeping data for one operational shift. As a result, the monthly amount of data is easier to maintain.

While it is fundamentally correct to store data with the time stamps provided by the CA server, this leads to problems when the clocks are not synchronized. It is naturally impossible to correlate data if the time stamps are wrong. After adding values stamped with inaccurate future dates, a channel becomes effectively unusable since the binary data format cannot insert values going back in time. The archive engine has constantly been enhanced to avoid these situations by neglecting erroneous time stamps like those which are too far ahead of the host computer clock, but ultimately the clocks have to be synchronized.

Most users seem to prefer the web interface (CGIExport) for data retrieval because they can access this remotely without having to install further software.

## 7 CONCLUSION

The Channel Archiver has proven to be a useful toolset for collecting and retrieving historical data. It is constantly being extended to meet new user needs, especially in the area of data retrieval. Scripting languages allow arbitrary correlation and other data mining operations, but since these generic languages are not specifically designed for numerical operations and data visualization, Matlab support was added.

Current plans include further optimization of the CGIExport tool because of its wide user acceptance. Matlab could benefit from native archive access, eliminating the intermediate step of exporting data into a file that is then parsed into Matlab. A common graphical user interface for the management of running ArchiveEngines, their configuration and periodic restart as well as maintenance of the binary archive data files would help new users.

*Work supported by the Office of Basic Energy Science, Office of Science of the US Department of Energy, and by Oak Ridge National Laboratory.*